\documentclass[aps,prb,twocolumn]{revtex4}
\usepackage{graphicx,amsmath,amssymb}
\usepackage[usenames,dvipsnames]{pstricks}
\usepackage{epsfig}

\begin{document}

\title{How and why to think about scattering in terms of wave packets \\
  instead of plane waves}
\author{Travis Norsen, Joshua Lande}
\affiliation{Marlboro College \\ Marlboro, VT  05344}
\author{S. B. McKagan}
\affiliation{JILA, University of Colorado and NIST \\ Boulder, CO, 80309}

\date{April 17, 2009}

\begin{abstract}
We discuss ``the plane wave approximation'' to quantum mechanical
scattering using simple one-dimensional examples.  Our central claims
are that (a) the plane waves of standard calculations can and should
be thought of as very wide wave packets, and (b) the calculations of
reflection and transmission probabilities $R$ and $T$ in standard
textbook presentations involve an approximation which is almost never
discussed.  We present a simple and intuitively revealing alternative 
way to derive and understand the connection between asymptotic wave
function amplitudes and scattering probabilities, which also has the
benefit of bringing the approximate character of standard plane wave
calculations out into the light.  We then develop an under-appreciated 
exact expression for scattering probabilities, using it to calculate, 
for two standard examples, expressions for $R$ and $T$ for an incident 
wave packet.  Comparing these results to the corresponding probabilities 
calculated using the plane wave approximation helps illuminate the
domain of applicability of that approximation and thus underscores the
importance of thinking about scattering in terms of wave packets
instead of plane waves. 
\end{abstract}

\maketitle

\section{Introduction}

Scattering is arguably the most important topic in quantum physics.
Virtually everything we know about the micro-structure of matter, we
know from scattering experiments.  And so the theoretical techniques
involved in predicting and explaining the results of these experiments
play a justifiably central role in quantum physics courses at all
levels in the physics curriculum, from Modern Physics for sophomores
through Quantum Field Theory for graduate students.

Given the importance of this topic, we should be
particularly careful about clarifying its physical and conceptual
foundations -- both for ourselves and for our students.
Unfortunately, the standard treatment of scattering (where the
scattering particle is described as a 
plane wave, rather than a propagating, normalizable,
finite-width wave packet) contains serious conceptual difficulties.
In Section \ref{sec2} we will briefly review the familiar plane wave 
approach to calculating reflection and transmission probabilities from 
a step potential and identify several conceptual problems that this
approach presents.

In Section \ref{sec3}, we present a novel alternative approach based
on a straightforward analysis of the kinematics of wave packets which 
we believe provides a superior conceptual foundation for thinking
about and introducing students to scattering.  We thus demonstrate 
that many of the conceptual problems associated with the plane wave 
analysis (as well as some otherwise-pointless mathematical
complications) can be quite simply
avoided -- all while preserving the mathematical simplicity and
accessibility of the standard plane wave calculation.

A crucial implication of the proposed alternative approach is that
standard plane wave calculations are \emph{approximations} to
realistic scattering events (where the relevant wave packet will
always have finite spatial support).
In order to illuminate this point, we illustrate, in the two
subsequent sections, the relation between plane wave
and wave packet treatments of scattering with two simple examples
intended to illustrate two qualitatively different ways in which the
plane wave analysis is imprecise or potentially misleading.
In Section \ref{sec4}, we analyze the scattering of a Gaussian
packet from a step potential, using this to develop a much more general
exact expression for reflection and transmission probabilities.
For this example, we show that the
exact probabilities can be expanded in
powers of the inverse packet width, with the individual terms
analytically calculable.  We thus show explicitly that the usual
plane wave expressions for $R$ and $T$ emerge only in the limit of
a very wide packet.   In Section \ref{sec5} we treat the
reflection and transmission of a packet from a rectangular barrier and
show that, in a certain limit, the plane wave treatment is not just
slightly off, but instead badly misrepresents the qualitative
behavior of the actual solution.

It may appear obvious to some readers that the standard plane wave
analysis of scattering involves an approximation or
idealization, and that, strictly speaking, scattering should always be
thought of in terms of finite-width wave packets.  
However, most textbooks do not treat it as such.
While some do discuss scattering in terms of wave packets
\cite{Cohen-Tannoudji1977a,Shankar1994a,French1978a,Eisberg1985a,Robinett1997a,Tipler2002a,Ohanian1995a,Griffiths2005a}
(and see also References \onlinecite{Hobbie1962},
\onlinecite{Gettys1965}, and Chapter 6 of \onlinecite{Newton1966a}),
these tend to be more advanced texts.  They also 
tend to present wave packets as a qualitative afterthought, with
little discussion of the connection between the wave packets and the
plane waves that are, in the end, always used for actual calculations.  
Textbooks also often present wave packets in a misleading way -- for
example, by implying that wave packets merely provide a more physical
and conceptually realistic way of re-deriving the really-correct plane wave
expressions for $R$ and $T$.~\cite{Cohen-Tannoudji1977a}  Further,
these texts never clarify the nature of the approximation involved in
the use of plane waves, nor do they address the domain of validity of
this approximation; indeed, nearly all
of these texts contain pictures of wave packets whose width is
not much larger than the average incident wavelength and which (as
we will show) are thus in fact very badly approximated by plane waves.

We note that in the vast majority of
experimentally realizable situations, the wave packets are so wide
(compared to other relevant length scales such as the de Broglie
wavelength) that treating them as infinitely wide, i.e., as plane
waves, is an excellent and appropriate approximation.  Our concern is 
thus not with the correctness of the plane wave approach, but with 
the underlying conceptual and pedagogical issues:  presenting
scattering from the very
beginning in terms of plane waves obfuscates what can and should be
the clear connections between the formal calculations and the physical
process of scattering, and hence pointlessly confuses students.  It
also obscures the fact that a plane wave treatment necessarily
involves at least some degree of approximation.
Hence our conclusion:  it is in terms of wave packets
that we should think about scattering ourselves, and introduce
scattering to students.

\section{The plane wave account and its problems}
\label{sec2}
Most students first encounter the quantum mechanical treatment of
scattering with the simple example of a 1-D particle incident on a
potential step:
\begin{equation}
V(x) = V_0 \; \theta(x) = \left\{
\begin{array}{lll}
0 & {\mathrm{for}} & x<0 \\
V_0 & {\mathrm{for}} & x>0 \\
\end{array}
 \right.
.
\label{step}
\end{equation}
We will base our discussion on this example, although everything we
will say can be applied to scattering
problems in general (including 3D problems, where $R$ and $T$ are
replaced by the differential cross section).

We begin this section with the familiar plane wave calculation of $R$
and $T$ probabilities for the potential step as it is presented in most
textbooks.~\cite{Shankar1994a,French1978a,Eisberg1985a,Robinett1997a,Tipler2002a,Beiser2003a,Harris1998a,Landau1977a,Goswami1997a}
We then discuss the possible conceptual 
problems this approach presents for students.  (See also
Ref. \onlinecite{McKagan2008d} for research on related student
difficulties with plane waves.)

The standard derivation begins by finding solutions to the time-independent Schr\"odinger equation
\begin{equation}
-\frac{\hbar^2}{2m} \; \frac{\partial^2}{\partial x^2}\psi(x) + V(x) \; \psi(x) = E \; \psi(x)
\label{tise}
\end{equation}
valid on the two sides of the origin:
\begin{equation}
\psi_k(x) = \left\{
\begin{array}{lll}
A e^{ikx} + B e^{-ikx} & {\mathrm{for}} & x<0 \\
C e^{i\kappa x}  & {\mathrm{for}} & x>0
\end{array}
\right.
\label{abc}
\end{equation}
where $k^2 = 2 m E/\hbar^2$ and $\kappa^2 = k^2 - 2 m V_0/\hbar^2$.
In principle the solution for $x>0$ should be supplemented by an
additional term:  $D e^{-i \kappa x}$.  This term is omitted
on the grounds that one is envisioning a particle
incident from the left (the $A$ term):  it may reflect back
toward the left (the
$B$ term) or transmit through to the right (the $C$ term), but cannot
be found on the right moving to the left.

Imposing the usual continuity conditions on $\psi$ at $x=0$ gives the following expressions relating the
amplitudes of the incident, reflected, and scattered waves:
\begin{equation}
\frac{B}{A} = \frac{k - \kappa}{k + \kappa}
\label{B}
\end{equation}
and
\begin{equation}
\frac{C}{A} = \frac{2k}{k + \kappa}.
\label{C}
\end{equation}
The reflection ($R$) and transmission ($T$) probabilities are then
given by
\begin{equation}
R =
\frac{\left|C\right| ^2}{\left|A\right|^2} = \left( \frac{k - \kappa}{k + \kappa} \right)^2
\label{R}
\end{equation}
and
\begin{equation}
T =
\frac{\kappa}{k}  \frac{\left|C\right| ^2}{\left|A\right|^2}
= \frac{4 k \kappa}{(k+\kappa)^2}.
\label{T}
\end{equation}
These expressions are often (though not exclusively)
justified by reference to the \emph{probability current}
\begin{equation}
j = \frac{-i \hbar}{2m}\left( \psi^* \frac{\partial \psi}{\partial x}
  - \psi \frac{\partial \psi^*}{\partial x} \right).
\label{j}
\end{equation}
which describes the flow of quantum mechanical probability.

For a wave or wave component with a plane wave structure, e.g.,
$\psi_A = A e^{ikx}$, Equation~\eqref{j} gives the
probability current
\begin{equation}
j_A = \frac{\hbar k}{m}|A|^2
\label{jA}
\end{equation}
and similarly for $j_B$ and $j_C$.
The plane wave probability
current can also be understood as the probability density ($|A|^2$)
multiplied by the group velocity
\begin{equation}
v_g(k) = \frac{d \omega(k)}{dk} = \frac{\hbar k}{m}
\label{vgroup}
\end{equation}
where $\omega(k) = E(k)/\hbar = \hbar k^2 / 2m$.

The reflection and transmission coefficients are given by the ratios of
the individual probability currents for the reflected and transmitted
terms to the incident current:
\begin{equation}
R = \frac{|j_R|}{j_I}
\label{rj}
\end{equation}
and
\begin{equation}
T = \frac{j_T}{j_I}.
\label{tj}
\end{equation}
Substituting Equation \eqref{jA} and the analogous expressions for $j_B$ and $j_C$ into Equations \eqref{rj}-\eqref{tj} yields Equations \eqref{R}-\eqref{T}.

While this approach appears concise, clear, and rigorously correct to
a physicist who is already familiar with the concepts involved, it may
be confusing to a student encountering scattering for the first time
for several reasons.

First and foremost, the basic method of using the general solution to
the time-independent Schr\"odinger equation is disconnected from the
time-dependent intuitive picture that both physicists and students use
to describe the physical situation of scattering: we say that
particles propagate from the left, reflect or transmit at $x=0$, and
subsequently propagate out to the left or right.  This description
assumes a very specific physical situation, namely that of a wave packet
approaching the step with some definite width, position, speed, and
time.  This physical picture is in play, tacitly, in the justification
of several of the
steps in the standard plane wave derivation.  For example, the fact
that particles \emph{propagate} suggests the choice of complex
exponentials rather than sines and cosines in Equation \eqref{abc},
and the fact that particles must begin in some specific location
suggests the elimination of the $D$ term.  However, most textbooks do
not make any explicit attempt to connect the plane wave expressions to
the intuitive time-dependent physical picture described here.
This invites a serious disconnect in the minds of students
between the formal development and the physical process the formalism
is supposed to be describing -- i.e., it works against efforts to
encourage students ``to think of the problem statement as describing
a physical process -- a movie of a region of space during a short time
interval...'' \cite{vanHeuvelen1991a}

A further conceptual problem is that certain aspects of Equations
\eqref{R}-\eqref{T} are not intuitively clear.  In particular, it is
difficult to give an intuitive explanation for the factor of
$\kappa/k$ which enters in Equation \eqref{T}.
According to the standard probability
interpretation of the wave function, $R$ and $T$ should be given by
the area under the reflected and transmitted parts of $|\psi|^2$,
respectively (assuming the wave function is normalized).  Since these
areas are infinite for plane waves, one can't calculate that as one
would naively expect.  It is quite tempting (and quite wrong) to
assume that the infinite widths simply cancel, giving $R=|B|^2/|A|^2$
and $T=|C|^2/|A|^2$ (without the factor of $\kappa/k$).  We have observed
that this is a common mistake among both physics students and teachers,
but most textbooks do not confront it directly.

Textbooks explain the equations for $R$ and $T$ using one of the
following approaches: (a) deriving Equations \eqref{R}-\eqref{T} using
probability
current~\cite{Shankar1994a,French1978a,Eisberg1985a,Robinett1997a,Tipler2002a,Beiser2003a,Harris1998a,Landau1977a,Goswami1997a}
as we sketched above,
(b) giving these equations without
explanation~\cite{Tipler2002a,Cohen-Tannoudji1977a}, (c) avoiding
these equations altogether by skipping the step potential and going
straight to tunneling through a square barrier, using $T=|C|^2/|A|^2$
without mentioning that this happens to be correct only for the
special case where the potential is equal on both sides of the
barrier~\cite{Griffiths2005a,Taylor2004a,Krane1996a,Serway2005a,Ohanian1995a,Knight2004a}
, or -- perhaps most bizarrely -- (d) stating that the factor of $k$ in
Equation \eqref{T} is due to ``accepted convention'' to ensure that
$R+T=1$~\cite{Eisberg1985a}.

Approaches (c) and (d) are misleading and encourage an incorrect
understanding that will lead students to errors when exposed to
scattering in new contexts.  Approach (b) does nothing to help
students understand the meaning of these equations or apply them in
other contexts.  Approach (a) is generally considered the most correct
and rigorous.  But even this approach introduces
difficulties for students.  Probably current is an important concept
that students need to learn, especially in an advanced course.
However, it is a rather sophisticated concept, and it is introduced in
the context of scattering, often for the first time, \emph{solely} for
the purpose of deriving $R$ and $T$.  Introducing such a concept in
the middle of a derivation places extra cognitive load on students,
increasing the likelihood that they will give up on understanding and
just accept the results ``on faith," as magic formulas to be memorized
and used without comprehension.  Further, it is difficult to make a
rigorous, rather than hand-waving, argument for why
Equations \eqref{rj}-\eqref{tj} are the correct expressions for $R$
and $T$ in terms of probability current.

One advantage of the probability current approach is that it does give
a somewhat intuitive explanation for the $\kappa$ and $k$ factors 
that appear
in Equation \eqref{T}: if the incoming and transmitted waves are
traveling at different speeds, then it makes sense that the amount
transmitted, and therefore the transmission probability, should be
proportional to the ratio of the speeds, which are in turn given by
Equation \eqref{vgroup}.  However, this advantage is
tempered by two problems.  First, this explanation is difficult to
relate to the standard probability interpretation and the intuition
that the probability should be given by the area under the curve of
$|\psi(x)|^2$.  Second, it is not intuitively clear why the relevant
speed should be the group velocity rather than the phase velocity.  
Indeed, it is not even clear what the group velocity means for a plane 
wave.~\footnote{The
  best and only way we know to think about group velocity for a plane
  wave is to imagine the plane wave as a very wide packet.  The group
  velocity is then the velocity at which the entire packet moves.} If
one tries to gain an intuitive understanding of the formula for the
transmission coefficient by looking at a simulation of plane waves
incident on a step
potential~\footnote{e.g. http://phet.colorado.edu/tunneling}, one will
actually be misled, since the only velocity that is apparent to the
eye, the phase velocity, has a response to changes in potential that
is opposite to that of the group velocity.    That is, in a 
region where the potential is increased (like
the $x > 0$ region of our step potential), the group velocity
\emph{decreases} whereas the phase velocity \emph{increases}.  The
attempt to intuitively understand Equations \eqref{R} - \eqref{T} 
as ratios of speeds-times-intensities for the various wave components 
is thus likely to fail.

In summary, the standard analysis of 1-D scattering in terms of plane 
waves, although mathematically simple, obscures the inherently
time-dependent physical nature of scattering, requires a
cognitive-load-increasing pedagogical detour through probability
current (or dubious and misleading hand-waving, or obfuscation),
and raises deep
questions which are not easily answerable without stepping back from
and getting underneath the plane wave approach.
Wouldn't it be nice if there were
some alternative approach that (a) didn't require the overhead
of probability current and (b) required
students to think, from the very beginning, that we are really dealing
with propagating, finite-width \emph{wave packets} to which the
plane waves are merely a convenient \emph{approximation}?  Just such an
approach will be presented in the following section.

\section{Scattering probabilities and packet widths}
\label{sec3}

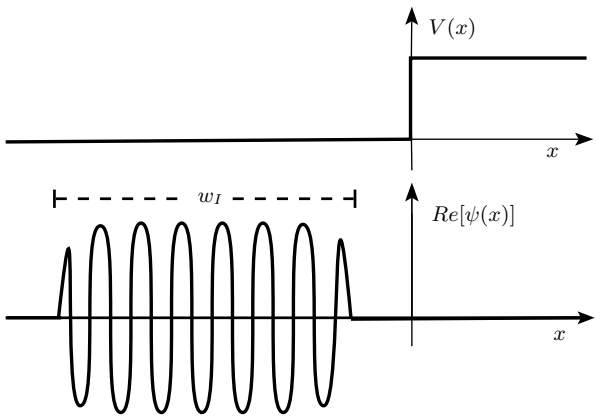
\begin{figure}[t]
\centering
\scalebox{0.9}{
% Generated with LaTeXDraw 1.9.5
% Fri Aug 01 14:23:36 EDT 2008
% \usepackage[usenames,dvipsnames]{pstricks}
% \usepackage{epsfig}
% \usepackage{pst-grad} % For gradients
% \usepackage{pst-plot} % For axes
\scalebox{1} % Change this value to rescale the drawing.
{
\begin{pspicture}(0,-3.025)(8.72,3.01)
\usefont{T1}{ptm}{m}{n}
\rput(6.601406,2.67){$V(x)$}
\usefont{T1}{ptm}{m}{n}
\rput(8.081407,0.85){$x$}
\psbezier[linewidth=0.05](5.1,-1.6)(4.8,1.6)(4.98,-2.888618)(4.7,-2.9)(4.42,-2.9113822)(4.7,-0.2886179)(4.4,-0.22)(4.1,-0.15138209)(4.4,-3.0)(4.1,-3.0)(3.8,-3.0)(4.1,-0.2)(3.8,-0.2)(3.5,-0.2)(3.8,-3.0)(3.5,-3.0)(3.2,-3.0)(3.5,-0.2)(3.2,-0.2)(2.9,-0.2)(3.2,-3.0)(2.9,-3.0)(2.6,-3.0)(2.9,-0.2)(2.6,-0.2)(2.3,-0.2)(2.6,-3.0)(2.3,-3.0)(2.0,-3.0)(2.3,-0.2)(2.0,-0.2)(1.7,-0.2)(2.0,-3.0)(1.7,-3.0)(1.4,-3.0)(1.7,-0.24)(1.4,-0.24)(1.1,-0.24)(1.3821588,-2.8975303)(1.1,-2.9)(0.81784123,-2.9024699)(1.1,1.3)(0.78,-1.6)
\usefont{T1}{ptm}{m}{n}
\rput(6.911406,-0.09){\psframebox*[framesep=0, boxsep=false,fillcolor=White] {$Re[\psi(x)]$}}
\psline[linewidth=0.04cm,linestyle=dashed,dash=0.16cm 0.16cm,tbarsize=0.07055555cm 5.0]{|-|*}(0.7,0.16)(5.16,0.16)
\psframe[linewidth=0.04,linecolor=White,dimen=outer,fillstyle=solid](2.94,0.42)(2.54,-0.08)
\usefont{T1}{ptm}{m}{n}
\rput(3.0714064,0.17){\psframebox*[framesep=0, boxsep=false,fillcolor=White] {$w_I \;$}}
\psline[linewidth=0.04cm,arrowsize=0.1529cm 1.0,arrowlength=1.31,arrowinset=0.2]{->}(0.040699005,-1.6)(8.7,-1.6)
\psline[linewidth=0.05](0.0,1.0)(5.98,1.04)(5.98,2.24)(8.58,2.24)
\psline[linewidth=0.02,arrowsize=0.1529cm 2.0,arrowlength=1.4,arrowinset=0.2]{<->}(6.0,3.0)(5.98,1.04)(8.68,1.04)
\psline[linewidth=0.024cm,arrowsize=0.1529cm 2.0,arrowlength=1.4,arrowinset=0.2]{->}(6.0,-1.6)(6.0,0.4)
\usefont{T1}{ptm}{m}{n}
\rput(8.181406,-1.85){$x$}
\psline[linewidth=0.05cm](0.0,-1.6)(0.82,-1.6)
\psline[linewidth=0.05cm](5.1,-1.62)(8.5,-1.6)
\psline[linewidth=0.02cm](5.98,2.04)(6.0,0.58)
\psline[linewidth=0.02cm](6.0,-0.06)(6.0,-2.02)
\end{pspicture} 
}
}
\caption{A generic wave packet (with approximately-constant amplitude
  over most of its width, $w_I$) is incident on the step potential's
  scattering center at $x=0$.}
\label{fig1}
\end{figure}

In this section we present an alternative derivation of the reflection
and transmission probabilities that uses wave packets instead of plane
waves.  We recommend basing students' first introduction to scattering
on this alternative approach.  It is simpler than the standard plane
wave derivation, builds on the standard $|\psi|^2$ definition of probability
(rather than the more sophisticated idea of probability current), and
provides a clear intuitive explanation for why the ratio of wave
numbers (or equivalently group velocities)
appears in the equation for the transmission coefficient.

Consider a wave packet approaching the scattering center at $x=0$ for
the potential defined in Equation \eqref{step}, as indicated in
Figure~\ref{fig1}.
Assume the packet has an almost-exactly constant amplitude ($A$)
and wavelength ($\lambda_0 = 2 \pi / k_0$)
in the region (of width $w_I$) where the amplitude is
non-vanishing, as
shown in the Figure.  Thus, where the amplitude is non-zero, the
packet will at each moment be well-approximated by a plane wave:
\begin{equation}
\psi = A \, e^{i k_0 x}.
\end{equation}
We may assume
this incident packet is normalized, so that $|A|^2 w_I = 1$.

What happens as the packet approaches and then interacts with the
potential step at $x=0$?  To begin with, the packet retains its
overall shape as it approaches the scattering center (that is, we
assume that the inevitable spreading of the wave packet is negligible
on the relevant timescales).  It simply moves at the group velocity
corresponding to the central wave number for the region $x<0$:
\begin{equation}
v_{g}^{<} = \frac{\hbar k_0}{m}.
\label{vg<}
\end{equation}
We then divide the scattering process into three stages:
\begin{itemize}
\item The leading edge of the packet arrives at $x=0$
\item The constant-amplitude ``middle'' of the packet is arriving at $x=0$
\item The trailing edge of the packet arrives at $x=0$
\end{itemize}
Suppose the leading edge arrives at time $t_1$.  Then the trailing edge will
arrive at $t_2$ satisfying
\begin{equation}
t_2 - t_1 = w_I / v_{g}^{<} = w_I m / \hbar k_0 .
\end{equation}
For intermediate
times, $t_1 < t < t_2$, we will have, in some (initially
small, then bigger, then small again) region surrounding $x=0$,
essentially the situation described in Equation \eqref{abc}, namely: a
superposition of rightward- and leftward-directed plane waves (just to
the left of $x=0$) and a rightward-directed plane wave with a
different wave number (to the right).  And the same
relations derived in the previous section for the relative amplitudes
of these three pieces will still apply.

While crashing into the scattering center, the incident packet
``spools out'' waves -- with amplitudes $B$ and $C$ given in
Equations \eqref{B} and \eqref{C} -- which propagate back to the left and
onward to the right, respectively.  These scattered waves will also be
wave packets, with their leading and trailing edges produced at times
$t_1$ and $t_2$ respectively.

This dynamical picture yields a very simple and illuminating 
alternative way to derive Equations \eqref{R}-\eqref{T}, that is, the
relations between the amplitude ratios ($B/A$ and $C/A$) and the
reflection and transmission probabilities.
Consider first the reflected packet.  The probability of reflection,
$R$, is by definition just its total integrated probability density --
which here will be its intensity $|B|^2$ times its width $w_R$.  But
the width of the reflected packet will be the same as the
width of the incident packet: because these two packets both
propagate in the same region, they have the same group velocity, so
the leading edge of the reflected packet will be
a distance $w_I$ to the left of $x=0$ when the trailing edge of the
reflected packet is formed. Thus, we have
\begin{equation}
R = w_R |B|^2 = w_I |B|^2 = \left| \frac{B}{A} \right|^2
\end{equation}
where we have used the normalization condition for the incident packet:
$w_I |A|^2 = 1$.

Similarly, the total probability associated with the transmitted wave
will be its intensity $|C|^2$ times its width $w_T$.  But $w_T$ will
be \emph{smaller} than $w_I$ because the group velocity on the right is
slower than on the left.  In particular: the leading edge of the
transmitted packet is created at $t_1$; the trailing edge is created
at $t_2$; and between these two times the leading edge
will be moving to the right at speed
\begin{equation}
v_{g}^{>} = \frac{\hbar \kappa_0}{m}
\end{equation}
where $\kappa_0^2 = k_0^2 - 2 m V_0 / \hbar^2$ is the (central) wave
number associated with the transmitted packet.  Thus, the width of the
transmitted packet -- the distance between its leading and trailing
edges -- is
\begin{equation}
w_T = v_{g}^{>} \; (t_2 - t_1) = \frac{\kappa_0}{k_0} w_I
\end{equation}
and so the transmission probability is
\begin{equation}
T = w_T |C|^2 = \frac{\kappa_0}{k_0} \left| \frac{C}{A} \right|^2
\end{equation}
in agreement with Equation \eqref{T}.

To summarize, one can derive the usual
plane wave expressions for $R$ and $T$
merely by considering the kinematics of wave packets, without ever
mentioning plane waves or probability current.  Further, the 
perhaps-puzzling
factor of $\kappa_0 / k_0$ in the expression for $T$ has an
intuitive and physically clear origin in the differing \emph{widths}
of the incident and transmitted packets, which in turn originates from
the differing group velocities on the two sides.

This route to the important formulas is both simpler
% (in that probability current need never be mentioned) 
and more physically
revealing than the one
traditionally taken in introductory quantum texts:  there is a clearly
defined initial condition and a definite process occurring in time;
probability only enters in the standard way (as an integral of the
probability density $|\psi|^2$); and the two quantities needed to
define the probabilities (the packet widths and amplitudes) are
arrived at separately and cleanly.

In addition, thinking
in terms of wave packets can help students recognize that the
formulas developed above for reflection and transmission probabilities
(and this point applies equally well to three-dimensional
scattering situations) are \emph{approximations} and to understand
when those approximations do and do not apply.

In particular, the argument presented here suggests that the
mathematical expressions for $R$ and $T$ above will apply only
in the limit of very wide incident packets.  This has several
aspects.  First, we are justified in neglecting the dynamical
spreading of the wave packet (and hence, e.g., treating the reflected
packet as having the same width as the incident packet) only if the
speed of spreading is very small compared to the group velocity, that is, if
$\Delta k \ll k_0$, where $\Delta k \sim 1 / \Delta x \sim 1 /
w_I$ is the width of the incident packet in k-space.  This implies
that $w_I \gg \lambda_0$:  the width of the wave
packet should be much larger than the characteristic wavelength.

Second, the plane wave style derivation of the amplitudes assumes
that, for some time interval (roughly, $t_1 < t < t_2$), the
wave function's structure in some (variable) spatial region around
$x=0$ is indeed given by Equation \eqref{abc}.  But these conditions
will simply fail to apply if the actual wave function is (in the
appropriate space and time regions) insufficiently plane wave-like,
e.g., if the amplitude of the wave varies appreciably over a length
scale $\lambda_0 = 2 \pi / k_0$.  Thus (assuming a smooth spatial
envelope for the packet) the formulas will be valid in the limit $w_I
\gg \lambda_0$, which is mathematically equivalent to the
limit noted previously.

Third, the statements made above about the group velocities for
the reflected and transmitted packets are not precisely true, because
the reflected and transmitted packets need not be precisely centered (in
$k$-space) about $k_0$ and $\kappa_0$ respectively.  This is because
the higher-$k$ components of the incident packet are (typically)
marginally more probable to transmit than the lower-$k$ components
(though the opposite behavior is also possible).  Hence, the
transmitted packet will (typically) peak around a value slightly
greater than $\kappa_0$, and the reflected packet will (typically)
peak around a value slightly less than $k_0$.  These changes, however,
will vanish for packets that have a narrow $k$-space distribution,
i.e., a large width in physical space.
(See Ref. \onlinecite{Bramhall1970} for an illuminating discussion of this
``velocity effect.'')

Note finally that for a scattering center which has some finite
width (e.g., a rectangular barrier), there
is an additional length scale in the problem, and qualitative
arguments also suggest that the plane wave type analysis should be
valid only in the limit where the packet width is large compared to
the spatial size of the scattering region:  basically, the qualitative
argument presented above (for the step potential, where scattering
only occurs at $x=0$) will go through unchanged only so long as the
width of the region where scattering occurs (e.g., the width of a
rectangular barrier) is much smaller than the packet width.

Even students who are just encountering quantum mechanical scattering
for the first time should be able to understand all of these points
(with the exception maybe of the third).
That is, they should be able to understand how
to think about scattering in terms of wave packets and how the
standard textbook formulas (derived using plane waves) should be
thought of as applying, as approximations, to wave packets that are
wide compared to other length scales in the problem, e.g., $\lambda_0$
and the width of the scattering region.

The relevant length scale
for the ``velocity effect'' mentioned above is the inverse of $dT_k/dk$,
where $T_k$ is the plane wave transmission probability for incident
wave number $k$.  Beginning students won't understand that.  But this
point overlaps with the more intuitively obvious point about the
scattering width, and so can be simply ignored.  There are also some
subtleties associated with the applicability of the group velocity
concept; for example, the length scale
over which the amplitude of the incident packet changes
appreciably (roughly the width of the packet's edges) should also be
(contrary to our Figure \ref{fig1}!)
large compared to the central de Broglie wavelength.

We thus acknowledge that, like the plane wave approach we criticize,
the approach outlined here has some subtle presuppositions which
students may not grasp and which may conceivably lead to confusion and
error.  Nevertheless, we think the wave packet kinematics approach
outlined here is far superior to the traditional derivations, in that
it is fundamentally built around the time-dependent physical process
of a scattering wave packet.  It thus clarifies and highlights the
essential physics, which the plane wave approach tends to obscure.

To emphasize the  claims made in the
preceding qualitative discussion,
we include in the next two sections two more exact
treatments of finite-width wave packets scattering from
some typical textbook potentials.  The main point is to concretize, 
with these two examples, the fact that the plane wave
approximation is valid only when the packet width is large compared to
all other physically relevant length scales.

\section{Gaussian wave packet scattering from a step potential}
\label{sec4}

It is possible to work out the \emph{exact} $R$ and
$T$ probabilities for a Gaussian wave packet incident on the
potential step of Equation \eqref{step}.  Most of the derivation
is worked out in several
texts. \cite{Cohen-Tannoudji1977a,Shankar1994a}  But
invariably these texts fail to write down the exact expressions for
$R$ and $T$ and instead make last-minute approximations which result
in the plane wave results developed previously.  But it is worth
pushing through the calculation to the end, if only to illustrate that
the end exists and that the results reduce to the plane wave formulas
\emph{only in the appropriate limits}.  Having the exact result in
hand also allows one
to analytically pick off explicit expressions for first non-vanishing
corrections to the plane wave result, which is a great example
calculation to share with students.  That the corrections are small
in precisely the limits discussed qualitatively at the end of the
previous section, is also a nice confirmation of that discussion.

We begin with an incident Gaussian wave packet, with central
wave number $k_0$ and width $\sigma$ and centered, at $t=0$, at $x = -a$:
\begin{equation}
\psi(x,0) = (\pi \sigma^2)^{-1/4} e^{i k_0 (x+a)} e^{-(x+a)^2/2\sigma^2}.
\end{equation}
We then follow Shankar's text\cite{Shankar1994a} and proceed in four steps.

Step 1 is to find appropriately normalized energy eigenfunctions for
the step potential.  These may be parametrized by $k$ and are (up to
normalization) just the plane wave states given previously:
\begin{multline}
\psi_k(x) = \frac{1}{\sqrt{2\pi}} \left[ \left( e^{ikx} + \frac{B}{A} \;
    e^{-i k x} \right) \theta(-x) \right. \\
    \left. + \frac{C}{A} \; e^{i\kappa x} \theta(x) \right]
\end{multline}
where, as before, $\kappa^2 = k^2 - 2mV_0/\hbar^2$ and
$B/A$ and $C/A$ are to be interpreted as the \emph{functions of $k$}
given by Equations \eqref{B} and \eqref{C}.  The overall factor of
$1/\sqrt{2\pi}$ out front is chosen so that
\begin{equation}
\int \psi_{k'}^*(x) \psi_k(x) dx = \delta(k - k').
\end{equation}
We are here assuming that only eigenstates with energy
eigenvalues $E = \hbar^2 k^2 / 2m > V_0$ will be present in the
Fourier decomposition of the incident packet and hence
we make no explicit special provision for those $\psi_k$ for
which $\kappa$ is imaginary.  (For a more rigorous treatment see Ref.
\onlinecite{Hammer1977}; the approximations we introduce here don't change
any of the central conclusions.)  Note also that there are two linearly
independent states for each $E$ only one of which is included here.
The orthogonal states will have incoming, rather than exclusively
outgoing, plane
waves for $x>0$; such states will never enter given our initial
conditions.

Step 2 is to write the incident packet as a linear combination of the
$\psi_k$s:
\begin{equation}
\psi(x,0) = \int \psi_k(x) \; \phi(k,0) \; dk
\end{equation}
where (assuming $\sigma \ll a$ so the amplitude of the incident packet
vanishes for $x>0$)
\begin{equation}
\phi(k,0) = \left( \frac{\sigma^2}{\pi}\right)^{1/4} e^{-(k-k_0)^2 \sigma^2 / 2}
e^{i k a}
\end{equation}
turns out to be the ordinary Fourier Transform of $\psi(x,0)$.

Step 3 is to write $\psi(x,t)$ by appending the time-dependent
phase factor to each of the energy eigenstate components of
$\psi(x,0)$:
\begin{align}
\psi(x,t) =& \int \psi_k(x) \; \phi(k,t) \; dk \nonumber \\
=& \int \psi_k(x) \;  \phi(k,0) \; e^{-i E(k) t / \hbar} \; dk
\nonumber \\
=& \left( \frac{\sigma^2}{4 \pi^3} \right)^{1/4} \int
e^{-i \hbar k^2 t / 2m}
e^{ - (k-k_0)^2 \sigma^2 / 2}
e^{i k a} \nonumber \\
& \times \left[ e^{ikx} \theta(-x) + \left( \frac{B}{A} \right)
 e^{-ikx} \theta(-x) \right. \nonumber \\
& \left. \; \; \; \; \;
 + \left( \frac{C}{A} \right) e^{i \kappa x}
\theta(x) \right] dk.
\label{wffull}
\end{align}

We can then finally -- Step 4 -- analyze the three terms for physical
content.  The first term, aside from the $\theta(-x)$, describes
the incident Gaussian packet propagating to the right.  For sufficiently
large times (when the incident packet would have support exclusively
in the region $x>0$) the $\theta(-x)$ kills this term -- i.e., the
incident packet eventually vanishes.

The second and third terms describe the reflected and transmitted
packets, respectively.  If the factors $(B/A)$ and $(C/A)$ were
constants, we would have Gaussian
integrals which we could evaluate explicitly to get exact expressions
for the reflected and transmitted packets -- which would themselves,
in turn, be Gaussian wave packets which could be (squared and) integrated
to get exact expressions for the $R$ and $T$ probabilities.  However,
these factors are functions of $k$.
It is not unreasonable to treat them as roughly constant
over the (remember, quite narrow) range of $k$ where $\phi(k,0)$ has
support.  This is the approach taken by Shankar (and, at least by
implication, several other texts) and the result is precisely the
plane wave expressions for $R$ and $T$ we developed earlier.

But another approach (which, surprisingly, we have not found in the
literature) is also appealing.  Consider the second and third terms
of Equation \eqref{wffull} -- which represent (for late times when these
terms are non-vanishing) the reflected and transmitted packets.
These can be
massaged to have the overall form (again assuming $t$ sufficiently
large that the $\theta$ factors can be dropped)
\begin{equation}
\psi_{R/T}(x,t) = \int \frac{e^{ikx}}{\sqrt{2\pi}} \; \phi_{R/T}(k,t) \; dk.
\end{equation}
Putting the two terms in this form requires a change of variables -- from
$k$ to $-k$ for the $R$ term, and from $k$ to
$\sqrt{k^2-2mV_0/\hbar^2}$ for the
$T$ term.  The resulting expressions for the $k$-space distributions
of the reflected and transmitted packets are:
\begin{multline}
\phi_R(k,t) = \\
\left( \frac{\sigma^2}{\pi} \right)^{1/4}
e^{ i \hbar k^2 t / 2m }
e^{- (k+k_0)^2 \sigma^2 / 2 }
e^{ -i k a }
\left(\frac{k+\kappa}{k-\kappa}\right)
\end{multline}
and
\begin{multline}
\phi_T(k,t) = \\
\left( \frac{\sigma^2}{\pi} \right)^{1/4}
e^{ -i \hbar (k^2+2mV_0/\hbar^2) t / 2m }
e^{ - (\sqrt{k^2+2mV_0/\hbar^2}-k_0)^2 \sigma^2 / 2 } \\
\times e^{i k a }
\left(\frac{2\sqrt{k^2+2mV_0/\hbar^2}}{\sqrt{k^2+2mV_0/\hbar^2}+k}\right)
\frac{k}{\sqrt{k^2+2mV_0/\hbar^2}}.
\end{multline}

To find the total probability associated with a given packet, we can
just as well integrate the momentum-space wave functions as the
position-space wave functions.  Thus,
\begin{align}
R &=\int \left| \phi_R(k,t) \right|^2 \; dk
\nonumber \\
&=\left( \frac{\sigma^2}{\pi} \right)^{1/2} \int
e^{-(k+k_0)^2 \sigma^2 } \left( \frac{k+\kappa}{k-\kappa} \right)^2 \; dk
\nonumber \\
&=\left( \frac{\sigma^2}{\pi} \right)^{1/2} \int
e^{-(k-k_0)^2 \sigma^2 } \left| \frac{B}{A}\right|^2 \; dk
\end{align}
where in the last step we have done another change of variables from
$k$ to $-k$.  This result can be summarized as follows:
\begin{equation}
R = \int P(k) R_k dk \label{int_for_R}
\end{equation}
where $P(k) = |\phi(k,0)|^2$ is the probability density for a given $k$
associated with the incident packet, and $R_k$ is simply the
reflection probability for a particular value of $k$ as expressed in
Equation \eqref{R}.

The analogous result for the $T$ term emerges after some more
convoluted algebra:
\begin{align}
   T =& \int \left| \phi_T(k,t) \right|^2 \; dk
   \nonumber \\
   =& \left( \frac{\sigma^2}{\pi} \right)^{1/2} \int
   e^{-(\sqrt{k^2+2mV_0/\hbar^2}-k_0)^2 \sigma^2}
   \nonumber \\
   & \; \; \; \; \; \; \; \; \; \;  \times
   \left( \frac{2\sqrt{k^2+2mV_0/\hbar^2}}{\sqrt{k^2+2mV_0/\hbar^2}+k} \right)^2
   \frac{k^2}{k^2+2mV_0/\hbar^2} dk
   \nonumber \\
   =& \left( \frac{\sigma^2}{\pi} \right)^{1/2} \int
   e^{ -(k-k_0)^2 \sigma^2 }
   \left| \frac{C}{A}\right|^2 \frac{\kappa}{k} \; dk \nonumber \\
=& \int P(k) T_k dk.\label{int_for_T}
\end{align}
where in the next-to-last step we have made a change of variables (back!)
from $k$ to $\sqrt{k^2+2mV_0/\hbar^2}$.

These expressions are exact (subject
to the assumptions noted earlier).  Note that, if we treat
$|B/A|^2$ and $(\kappa/k)|C/A|^2$ as constants
that do not depend on $k$ (i.e., if we approximate these functions by
their values at $k = k_0$, which is a good approximation so
long as as the functions don't vary appreciably in a region of width
$1/\sigma$ around $k_0$, i.e., if the width $\sigma$ of the incident packet is
very big) we are left with plain Gaussian integrals that can be evaluated
to get back the plane wave-approximation results we started with:
$R = |B/A|^2$ evaluated at $k = k_0$, etc.

Unfortunately, the actual integrals are too messy to do exactly.
But we can Taylor expand the complicating
factors around $k=k_0$ to get a series of standard integrals,
resulting in a power-series expansion (in inverse powers of the packet
width $\sigma$) for the exact $R$ and $T$.

The first two non-vanishing terms for $R$ and $T$ are as follows:
\begin{equation}
    R=\left(\frac{k_0-\kappa_0}{k_0+\kappa_0}\right)^2+
    \left(\frac{2k_0}{\kappa_0^3}+\frac{8}{\kappa_0^2}\right)
    \left(\frac{k_0-\kappa_0}{k_0+\kappa_0}\right)^2\frac{1}{\sigma^2} + \cdots
    \label{R_expand}
\end{equation}
and
\begin{equation}
    T= \frac{4 k_0 \kappa_0}{(k_0+\kappa)^2}
    - \left( \frac{2k_0}{\kappa_0^3} + \frac{8}{\kappa_0^2} \right)
    \left( \frac{k_0-\kappa_0}{k_0+\kappa_0} \right)^2
    \frac{1}{\sigma^2} + \cdots
    \label{T_expand}
\end{equation}
The first terms are just the standard plane wave results.
The leading-order corrections vanish (roughly) in the limit $\sigma^2
k_0^2 >> 1$, which confirms the conclusion of the qualitative
discussion in Section \ref{sec3}:  the plane wave approximation will
be accurate only if the packet width is very large compared to the de
Broglie wavelength, $\lambda_0 = 2 \pi / k_0$.

Note also that Equations \eqref{int_for_R}-\eqref{int_for_T}
hold much more generally than just for Gaussian
wave packets and step potentials.  One can assume an arbitrary
initial wave packet with $k$-space distribution $\phi(k,0)$
and arbitrary
potential function $V(x)$ (subject to the assumption that it is
asymptotically constant on both sides of the scattering region),
and still carry through the above derivation.  One cannot say much in
general about $R_k$ and $T_k$ for an arbitrary
potential, but Equations \eqref{int_for_R} - \eqref{int_for_T} will
still hold.

\section{Scattering from a Rectangular Barrier}
\label{sec5}

In the previous section we showed that, for scattering from a step
potential, the standard textbook formulas are approximations which
are increasingly correct as the width of the incident packet is taken
to infinity.  We now turn to a different example -- scattering from a
rectangular barrier -- to show more explicitly something else we claimed
earlier:  if the scattering region has a finite width that is not
small compared to the width of the incident packet, the plane wave
approximation will give badly misleading, qualitatively wrong results.

\begin{figure}[t]
\centering
\scalebox{0.9}{
% Generated with LaTeXDraw 1.9.5
% Fri Aug 01 14:26:12 EDT 2008
% \usepackage[usenames,dvipsnames]{pstricks}
% \usepackage{epsfig}
% \usepackage{pst-grad} % For gradients
% \usepackage{pst-plot} % For axes
\scalebox{1} % Change this value to rescale the drawing.
{
\begin{pspicture}(0,-3.87)(8.475262,3.88)
\psbezier[linewidth=0.04,fillstyle=solid](0.862274,-1.3070132)(1.2824361,-1.3139375)(1.1703929,-0.85)(1.3477947,-0.85)(1.5251964,-0.85)(1.4691749,-1.3139373)(1.88,-1.3139373)
\psbezier[linewidth=0.04,fillstyle=solid](0.47239915,1.04442)(1.0205778,1.0332125)(0.87439686,1.784115)(1.1058501,1.784115)(1.3373033,1.784115)(1.2642127,1.0332125)(1.8002096,1.0332125)
\psbezier[linewidth=0.04,fillstyle=solid](6.636153,-1.3204043)(7.0011034,-1.3183694)(6.897032,-0.9760568)(7.0510793,-0.9730284)(7.2051263,-0.97)(7.1632586,-1.3551816)(7.52,-1.29)
\psframe[linewidth=0.04,linecolor=White,dimen=outer,fillstyle=solid](3.756199,0.24608503)(3.0496576,0.03314254)
\psbezier[linewidth=0.04,fillstyle=solid](2.3874795,-0.124393426)(2.7972357,-0.12945999)(2.6879675,0.21)(2.8609755,0.21)(3.0339837,0.21)(2.9793496,-0.12945998)(3.38,-0.12945998)
\psbezier[linewidth=0.04,fillstyle=solid](1.28,-0.10977124)(1.7181509,-0.11825261)(1.6013106,0.45)(1.7863076,0.45)(1.9713047,0.45)(1.9128845,-0.118252374)(2.3412986,-0.118252374)
\psline[linewidth=0.04cm,arrowsize=0.1529cm 2.0,arrowlength=1.4,arrowinset=0.1]{->}(1.2885762,1.51272)(2.0072994,1.5239276)
\psline[linewidth=0.04cm,arrowsize=0.1529cm 2.0,arrowlength=1.4,arrowinset=0.1]{->}(3.0871115,0.14763254)(3.647472,0.14763254)
\psbezier[linewidth=0.04,fillstyle=solid](0.48227397,-2.4667146)(0.9024361,-2.4739375)(0.7903929,-1.99)(0.96779466,-1.99)(1.1451964,-1.99)(1.0891749,-2.4739373)(1.5,-2.4739373)
\psbezier[linewidth=0.04,fillstyle=solid](1.78,-2.4807398)(2.0692687,-2.485145)(1.9921304,-2.19)(2.1142662,-2.19)(2.2364018,-2.19)(2.1978326,-2.4851449)(2.480673,-2.4851449)
\psbezier[linewidth=0.04,fillstyle=solid](7.0655828,-2.4625218)(7.4430947,-2.46756)(7.342425,-2.13)(7.5018187,-2.13)(7.6612124,-2.13)(7.6108775,-2.4675598)(7.98,-2.4675598)
\psbezier[linewidth=0.04,fillstyle=solid](0.07612594,-3.6898303)(0.49056935,-3.6967976)(0.3800511,-3.23)(0.55503833,-3.23)(0.73002553,-3.23)(0.6747664,-3.6967974)(1.08,-3.6967974)
\psbezier[linewidth=0.04,fillstyle=solid](1.4990436,-3.7038558)(1.7966862,-3.708005)(1.7173148,-3.43)(1.8429861,-3.43)(1.9686574,-3.43)(1.9289718,-3.708005)(2.22,-3.708005)
\psbezier[linewidth=0.04,fillstyle=solid](6.5209155,-3.7157915)(6.7847576,-3.7192125)(6.7144,-3.533824)(6.8088818,-3.5119119)(6.903364,-3.49)(6.902021,-3.7192125)(7.16,-3.7192125)
\psbezier[linewidth=0.04,fillstyle=solid](7.366814,-3.7056375)(7.7438173,-3.71042)(7.6432834,-3.39)(7.8024626,-3.39)(7.961642,-3.39)(7.9113746,-3.71042)(8.28,-3.71042)
\psframe[linewidth=0.04,linecolor=White,dimen=outer,fillstyle=solid](3.44,-2.21)(2.910489,-2.41)
\psline[linewidth=0.04cm,arrowsize=0.1529cm 2.0,arrowlength=1.4,arrowinset=0.1]{->}(2.94,-2.31)(3.5,-2.31)
\psbezier[linewidth=0.04,fillstyle=solid](2.4684916,-2.4813368)(2.6961784,-2.485145)(2.6354618,-2.29)(2.74,-2.31)(2.844538,-2.33)(2.7973726,-2.4851449)(3.02,-2.4851449)
\psframe[linewidth=0.04,linecolor=White,dimen=outer,fillstyle=solid](6.2785125,-3.48627)(5.4617524,-3.69)
\psline[linewidth=0.04cm,arrowsize=0.1529cm 2.0,arrowlength=1.4,arrowinset=0.1]{->}(6.0749245,-3.57)(5.391784,-3.5671375)
\psline[linewidth=0.04cm,arrowsize=0.1529cm 2.0,arrowlength=1.4,arrowinset=0.1]{->}(5.736643,-1.0662475)(5.14,-1.07)
\psbezier[linewidth=0.04,fillstyle=solid](5.5,-1.3136477)(5.805505,-1.3176425)(5.7240367,-1.05)(5.8530273,-1.05)(5.9820185,-1.05)(5.941284,-1.3176425)(6.24,-1.3176425)
\psbezier[linewidth=0.04,fillstyle=solid](5.96,-3.7043128)(6.1994495,-3.7075598)(6.1355963,-3.55)(6.24,-3.57)(6.3444037,-3.59)(6.3058715,-3.7075598)(6.54,-3.7075598)
\psline[linewidth=0.04](2.4238718,2.2932127)(2.4238718,3.27)(6.460102,3.27)(6.460102,2.2932124)
\psline[linewidth=0.02cm,arrowsize=0.05291667cm 2.0,arrowlength=1.4,arrowinset=0.4]{->}(0.24293171,2.29)(8.369988,2.29)
\psline[linewidth=0.02cm,arrowsize=0.05291667cm 2.0,arrowlength=1.4,arrowinset=0.4]{<-}(2.42,3.87)(2.42,2.35)
\usefont{T1}{ptm}{m}{n}
\rput(2.9414062,3.64){$V(x)$}
\psframe[linewidth=0.0020,linecolor=White,dimen=outer,fillstyle=solid](8.2,1.05)(0.1,0.87)
\psframe[linewidth=0.0020,linecolor=White,dimen=outer,fillstyle=solid](8.14,-0.09)(0.04,-0.27)
\psframe[linewidth=0.0020,linecolor=White,dimen=outer,fillstyle=solid](8.2,-2.45)(0.1,-2.63)
\psframe[linewidth=0.0020,linecolor=White,dimen=outer,fillstyle=solid](8.28,-3.69)(0.0,-3.87)
\psframe[linewidth=0.0020,linecolor=White,dimen=outer,fillstyle=solid](8.22,-1.29)(0.12,-1.47)
\psline[linewidth=0.02cm,arrowsize=0.05291667cm 2.0,arrowlength=1.4,arrowinset=0.4]{->}(0.12,-3.67)(8.437648,-3.69)
\psline[linewidth=0.02cm,arrowsize=0.05291667cm 2.0,arrowlength=1.4,arrowinset=0.4]{<-}(2.42,2.01)(2.42,-3.69)
\psline[linewidth=0.0139999995cm,linestyle=dashed,dash=0.16cm 0.16cm](6.46,2.21)(6.46,-3.69)
\usefont{T1}{ptm}{m}{n}
\rput(3.0514061,1.62){$|\psi(x)|$}
\psline[linewidth=0.02cm,arrowsize=0.05291667cm 2.0,arrowlength=1.4,arrowinset=0.4]{->}(0.20293172,1.05)(8.3299885,1.05)
\psline[linewidth=0.02cm,arrowsize=0.05291667cm 2.0,arrowlength=1.4,arrowinset=0.4]{->}(0.1767154,-0.09)(8.34,-0.09)
\psline[linewidth=0.02cm,arrowsize=0.05291667cm 2.0,arrowlength=1.4,arrowinset=0.4]{->}(0.12428276,-1.29)(8.3824215,-1.29)
\psline[linewidth=0.02cm,arrowsize=0.05291667cm 2.0,arrowlength=1.4,arrowinset=0.4]{->}(0.20712265,-2.45)(8.465261,-2.45)
\end{pspicture} 
}
}
\caption{The top frame shows $V(x)$ for the rectangular barrier.
  Subsequent frames show how $|\psi(x)|$ evolves in time for a wave packet
  incident from the left, through an infinite sequence of
  back-and-forth reflections inside the barrier.  Note that the
  sizes of reflected packets are exaggerated relative to transmitted
  packets, compared to the assumptions made in the text.}
\label{fig2}
\end{figure}
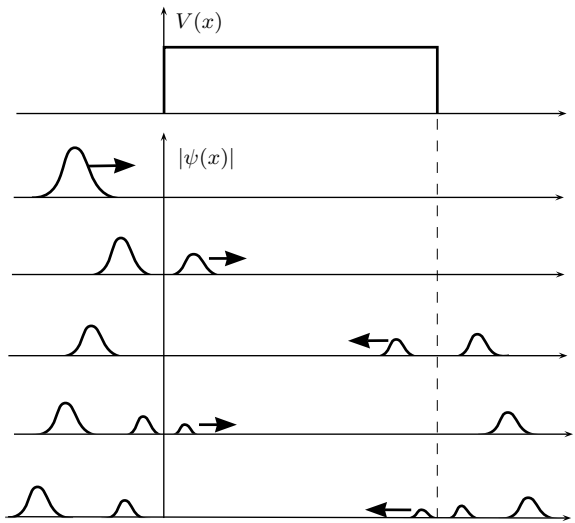

Consider the potential barrier given by
\begin{equation}
\label{rectangular_potential_equation}
        V(x)=
        \begin{cases}
        0   & x<0 \\
        V_0 & 0 \leq x \leq a \\
        0  &  x>0 .
        \end{cases}
\end{equation}
For a packet that is narrow compared to the barrier width $a$, but
still wide enough that we can ignore its dynamical spreading during
the time interval of the collision, there is an elegant kinematical
argument that allows one to work out the total reflection and
transmission probabilities.  See Figure~\ref{fig2}.
Following the principle (sometimes attributed to Wheeler) of never
calculating anything until one already knows the result, we sketch
this argument first.

For this argument, and the calculations that follow, we make several
assumptions.  First, we assume that the barrier is much
wider than the wave packet, that is, $a \gg w$ or $\Delta k \gg 1/a$,
where $w \sim 1/\Delta k$ is the width of the packet.  This
approximation ensures that we can treat the interaction of the packet
with each side of the barrier independently.  Second, we assume that
the wave packet is wide compared to its central wavelength $\lambda_0
= 2\pi/k_0$, that is, $w \gg \lambda_0$ or $k_0 \gg \Delta k$.  This
ensures that the packet does not spread out too much,
and warrants making a plane wave approximation in treating the
interaction of the packet with each edge of the barrier.  Finally, we
assume that $k_0 \gg \sqrt{2 m V_0 / \hbar^2}$, or $E \gg V_0$.
This is not strictly necessary, but allows us to simplify the
calculations by taking the reflection coefficient to be small.

Here's the kinematical argument.
When the packet arrives at the left side of the barrier, there is
probability $R_{k_0}$ for it to reflect, where the $R_{k_0}$ here
(and the $T_{k_0}$ to follow) is the
probability for reflection (transmission) \emph{from a step}.
The packet may also transmit, then reflect off the step on
the far side of the barrier, then transmit back out to the left:  this
has overall probability $T_{k_0}R_{k_0}T_{k_0}$.  Similarly, it could
reflect 3, 5, or any higher odd number of times inside the barrier
before finally leaving to the left.
(Note that $R_{k_0}$ and $T_{k_0}$
are the same whether one is going ``up'' or ``down'' the step.)
The total probability of reflection
is equal to the sum of all these possibilities (which are
non-interfering so long as the packet remains narrow compared to the
distances between adjacent packets):
\begin{align}
    R_{k_0}^{\text{(total)}} &= R_{k_0}+T_{k_0}R_{k_0}T_{k_0}+
    T_{k_0}R_{k_0}^3T_{k_0}+ \cdots \nonumber \\
    &= 2R_{k_0}/(1+R_{k_0}) \label{cute_argument_equation}
\end{align}
where we have summed a geometric series and
used the fact that $R_{k_0} + T_{k_0} = 1$.

In the limit that $E \gg V_0$, we have $R_{k_0} \ll 1$, and Equation~\eqref{cute_argument_equation} reduces to
\begin{equation}
R_{k_0}^{\text{(total)}} = 2R_{k_0}. \label{R_kinematic}
\end{equation}

What happens if we instead
calculate the reflection probability in the standard textbook way --
namely, using the energy eigenstates of the Schr\"odinger equation to
read off the reflection probability according to $R = |B/A|^2$?  That
is, what if we use the plane wave approximation?
We will not repeat this calculation, as it can be found in many
quantum mechanics textbooks, \cite{Cohen-Tannoudji1977a} but simply
quote the result for the total reflection probability for a barrier
potential:
\begin{align}
    R_{k_0}^{(\text{total})} = \left| \frac{B}{A} \right|^2
    & = \frac{(k_0^2-\kappa_0^2)^2\sin^2(\kappa_0 a)}{4k_0^2\kappa_0^2
    +(k_0^2-\kappa_0^2)\sin^2(\kappa_0 a)}
     \label{R_for_rectangle_barrier}
\end{align}
where $A$ and $B$ are the amplitudes of the incident and reflected
plane waves, respectively.
Rewriting in terms of the reflection and transmission coefficients for the
step potential, $R_{k_0}$ and $T_{k_0}$, from Equations \eqref{R} and
\eqref{T}, gives:
\begin{equation}
R_{k_0}^{(\text{total})} = \frac{4R_{k_0}\sin^2(\kappa_0
  a)/T_{k_0}^2}{1+4R_{k_0}\sin^2(\kappa_0 a)/T_{k_0}^2}.
\end{equation}
As before, because $R_{k_0} \ll 1$, we may simplify this to:
\begin{equation}
R_{k_0}^{(\text{total})} = 4 R_{k_0} \sin^2(\kappa_0 a)
\label{R_planewave}
\end{equation}
where, just to be clear, the left-hand-side refers to the overall
reflection probability from the rectangular barrier and the $R_{k_0}$ on
the right hand side denotes the probability for reflection from a
corresponding step potential, which is convenient since we want to
compare with Equation \eqref{R_kinematic}.

This is the standard plane-wave approximation for the $k_0$-dependent
reflection probability from a rectangular barrier (for large $E$).
Note in particular
that -- unlike the result of the kinematical argument sketched above
-- it oscillates rapidly back and forth between zero and $4 R_{k_0}$
as $k_0$ varies, with a period of order $1/a$.  So, if we believe the
kinematical argument, we can already see a major qualitative
disagreement between the reflection probability given by Equation~\eqref{R_kinematic}, calculated using
wave packets, and Equation~\eqref{R_planewave}, calculated from the plane-wave approximation.

The rigorous way to calculate the reflection probability for our wave
packet is to use
Equation \eqref{int_for_R}.  Part of the setup here was the assumption
that the packet width was small compared to the width of the barrier,
i.e., that $\Delta k$ -- the width of the packet in $k$-space -- is
large compared to $1/a$.  This implies that the plane-wave result,
Equation \eqref{R_planewave}
will oscillate back and forth between its minimum and maximum
values \emph{many times} over the support of $P(k)$.  And so the total
reflection probability will simply be the average value of the
plane-wave result:
\begin{equation}
R_{k_0}^{\text{(total)}} = 2 R_{k_0}
\end{equation}
in agreement with Equation~\eqref{R_kinematic}, the result of the
kinematical argument.

Whereas the plane wave approximation predicts that the reflection
probability rapidly oscillates between 0 and $4R_{k_0}$,
the actual probability is just $2R_{k_0}$ --  insensitive to the
precise value of $k_0$.
(The related oscillatory behavior of $R$ as the width
of the barrier is varied is displayed in Figure 10 of
Ref.~\onlinecite{Bramhall1970}, which treats scattering in terms of
wave
packets.  However, this figure shows only barrier widths that are
smaller than the packet width.  Yet, interestingly, it is suggested
by the figure that the oscillations disappear as the barrier width
approaches the packet width, which is consistent with our claims here.)

In addition to showing how the plane-wave approximation can give
qualitatively wrong results when it is applied inappropriately,
this example helps illustrate the useful generality
of Equation \eqref{int_for_R} for the
reflection coefficient. As the earlier step potential example showed,
this formula
correctly predicts the behavior in situations where the incoming
wave packet is wide compared to the barrier (i.e., it reduces to the
plane wave result in the appropriate limit).  But as the
rectangular barrier example shows, it also correctly predicts the
behavior in situations where the incoming wave packet is small in
comparison to some other length scale in the problem such that the
plane wave approximation badly fails to capture the true, qualitative
behavior of $R$.

\section{Conclusions}

The standard introductory textbook presentations of quantum mechanical
scattering are almost always in terms of unphysical plane wave states.
After reviewing some of the conceptual problems and mathematical
overhead of the plane wave approach, we showed in Section III
how these problems can be largely avoided
by instead deriving the basic expressions for reflection and
transmission probabilities from a simple analysis of the kinematics of
wave packets.  This allows students to genuinely understand the
connection between the formal calculations and the physical process of
scattering, and hence encourages visualization and physical thinking (e.g.,
about relative length scales in the problem).
In addition, the several (overlapping) approximations which enter
into the analysis can be understood intuitively even by beginning
students, which should help them understand that the $R$ and $T$
coefficients they encounter in textbooks (and homework problems) are
\emph{approximations} which are valid only in the limit of very wide
packets.

To highlight the approximate character of the standard plane wave
results, we computed exact expressions for $R$ and $T$ for a
Gaussian wave packet incident on a step potential.  We showed that
these can be written in the suggestive (and, as it turns out, general)
form of Equations \eqref{int_for_R} - \eqref{int_for_T}.  In the limit
of very wide packets, these reduce to the plane wave results; close to
that limit, it is possible and illuminating to use these formulas to
generate a power series in the inverse packet width.  This underscores
the point that the plane wave approximation can be -- and, in virtually
all actual experimental situations, \emph{is} -- very accurate.

We then briefly discussed reflection and transmission from a
rectangular barrier, such that the packet width is large compared to
the central wavelength, but small compared to the width of the
barrier.  We showed explicitly that the plane wave approximation badly
mischaracterizes the dependence of $R$ and $T$ on the central 
wave number $k_0$, whereas Equations \eqref{int_for_R} - \eqref{int_for_T}
handle everything automatically and correctly.  This illustrates the
general fact that the validity of the plane wave approximation relies
on the packet width being large, not only compared to the central de
Broglie wavelength, but large also compared to other length scales in
the problem.

In addition to the pedagogical value of introducing scattering in
terms of wave packets from the very beginning, our main conclusion is
the importance and fundamentality of Equations \eqref{int_for_R} -
\eqref{int_for_T}, which we have never seen in any undergraduate
quantum mechanics
texts and have in fact seen only very rarely in the literature.
(See, for example, References \onlinecite{Bramhall1970} and
\onlinecite{Garrido2008} and further references therein.
Ref. \onlinecite{Hobbie1962} seems to mention the result in words
without writing out the corresponding equations.)

It seems likely that the reason for this is that most physicists would
regard these expressions as \emph{obvious}.  An incident wave packet
is, after all, merely a superposition of plane wave states with some
associated wave number probability distribution $P(k)$, and the
reflection and transmission probabilities $R_k$ and $T_k$
can be defined individually for those states.  So it seems almost
irresistible to conclude that  Equations \eqref{int_for_R} - \eqref{int_for_T}
should be the right expressions.

But we don't think it is so obvious.  It cannot be so
quickly taken for granted that ``the reflection and transmission
probabilities $R_k$ and $T_k$ can be defined'' for the plane wave
states.  As we have discussed at length in Section \ref{sec2},
standard derivations of $R_k$ and $T_k$ are really inconsistent, in
that they presuppose a physical process that simply is not described
by the posited mathematics.  Simply put, plane waves do not
reflect or transmit.  The coefficients $R_k$ and $T_k$ are really
only meaningful if thought of the way we've advocated here -- as
probabilities for the reflection and transmission of wave packets, in
the limit where the packet width approaches infinity.

In short, it is the plane wave expressions that conceptually
presuppose wave
packets, not vice versa.  It is thus a mistake to think that
Equations \eqref{int_for_R} - \eqref{int_for_T} are trivially derivable
from more basic, independently meaningful things.  If anything, it
is a surprise that the probabilities can be written in this form.  But
they can, and this should be more widely known.  As should the general
perspective which gave rise to them:  thinking about scattering in
terms of wave packets.

\section*{Acknowledgements:}
Thanks to Mike Dubson and two anonymous referees for a number of
helpful comments on earlier drafts.

\bibliography{PlaneWaveFinal}
\bibliographystyle{apsrevtitlequotes}

\end{document}